\documentclass[aps,pra,superscriptaddress,onecolumn]{revtex4}
\usepackage{amsmath}
\begin{document}

\title {Comment on ``Force-field functor theory'' [arXiv:1306.4332]}

\author{R. Vaia}
\affiliation{Istituto dei Sistemi Complessi,
       Consiglio Nazionale delle Ricerche,
       via Madonna del Piano 10,
       I-50019 Sesto Fiorentino (FI), Italy}
       \affiliation{INFN Sezione di Firenze, via G.Sansone 1,
       I-50019 Sesto Fiorentino (FI), Italy}


\maketitle

This comment regards a recently published preprint by Babbush,
Parkhill and Aspuru-Guzik (henceforth BPA)~\cite{BabbushPA2013}. The
subject is the equilibrium thermodynamics of a system of many quantum
particles with Hamiltonian
 $\hat{H}=\sum_{i=1}^N\hat{p}_i^2/(2m)+V(\hat{q})$~,
where $\hat{q}=\{\hat{q}_i\}$ and the commutator of the coordinate-
and momentum operators is $[\hat{q}_i,\hat{p}_j]=i\hbar\delta_{ij}$.
In BPA it is correctly observed (see, e.g.,
\cite{HilleryCSW1984,CGTVV1995}) that the coordinate distribution
function at temperature $T=\beta^{-1}$, namely
\vspace{-2mm}\begin{equation}
 \eta(q)=\frac1Z~\langle{q}|e^{-\beta{\hat{H}}}|{q}\rangle
        \equiv \frac1Z~\bigg(\frac{m}{2\pi\hbar^2\beta}\bigg)^{N/2}
        e^{-\beta W(q)} ~,
\end{equation}
can be used to define the {\em effective} potential $W(q)$ (the
kinetic part of the partition function $Z$ is omitted in BPA) in
terms of which the exact quantum equilibrium average of any operator
$O(\hat{q})$ takes the classical form of a configuration integral,
\vspace{-2mm}\begin{equation}
 \big\langle O(\hat{q})\big\rangle
 = \int dq~\langle{q}|O(\hat{q})~e^{-\beta \hat{H}}|{q}\rangle
 = \frac1Z \bigg(\frac{m}{2\pi\hbar^2\beta}\bigg)^{N/2}
   \int dq~O(q)~e^{-\beta W(q)} ~.
\label{e.aveO}
\end{equation}
It is convincingly proven in BPA that the mapping
$\eta(q)\leftrightarrow{W}(q)$ (i.e., the exponential function) is
bijective, as well as $V(q)\leftrightarrow\eta(q)$, and then it
follows that $V(q)\leftrightarrow{W}(q)$ is one-to-one. Furthermore,
it is correctly pointed out that the
Giachetti-Tognetti-Feynman-Kleinert~\cite{GiachettiT1985,FeynmanK1986}
(GTFK) effective potential $V_{\rm{eff}}(q)$ differs from $W(q)$.
Indeed, $V_{\rm{eff}}(q)$, which accounts {\em exactly} for any
quadratic potential, entails that for approximating
$\langle{O}(\hat{q})\rangle$ one has to include a further Gaussian
average accounting for purely-quantum fluctuations, as shown, e.g.,
in \cite{CGTVV1995,VaiaT1990}: there $W(q)$ is also introduced and
dubbed the {\em local} effective potential. However, in BPA it is not
shown that Eq.~(29), the preprint's main result obtained as the
Jensen's approximation $W(q)\,{\approx}\,W_{_{\rm{BPA}}}(q)$ to the
exact formula~(26), can be calculated explicitly as
\begin{equation}
 W_{_{\rm{BPA}}}(q) =
  V(q)-\frac1\beta\big\langle{U[r(\tau)]}\big\rangle =
 \bigg\langle \int_0^{\beta\hbar}
        \frac{d\tau}{\beta\hbar}~V[r(\tau)]\bigg\rangle
 = \int \frac{d^N\xi}{(2\pi\sigma^2)^{\frac N2}}
     ~V(q{+}\xi)~e^{-\frac{\xi^2}{2\sigma^2}}~,
\label{e.Wappr}
\end{equation}
i.e., the convolution between the potential $V(q)$ and a Gaussian
with variance $\sigma^2={\beta\hbar^2}/(6m)$ proportional to the
squared de-Broglie wavelength. This is in agreement with the Wigner
series~\cite{Wigner1932} up to lowest order, but lacks the nonlinear
contributions to it. How accurate is the approximation made in
Eq.~(29) of BPA? One can estimate this by considering a single
($N{=}1$) quantum harmonic oscillator,
$V(\hat{q})\,{=}\,\kappa\,\hat{q}^2/2$, whose frequency is
$\omega\,{\equiv}\,\sqrt{\kappa/m}$. Expanding $V(q{+}\xi)$ in
Eq.~\eqref{e.Wappr} one finds
\begin{equation}
 W_{_{\rm{BPA}}}(q) = \sum_{n=0}^\infty
 \frac1{n!}~\frac{d^{2n}V(q)}{dq^{2n}}~\Big(\frac{\beta\hbar^2}{12m}\Big)^n
 = V(q) + \frac{\beta(\hbar\omega)^2}{12} ~,
\label{e.ho.Wappr}
\end{equation}
which does not improve upon the classical result using
Eq.~\eqref{e.aveO}. From the known density for the quantum harmonic
oscillator
the {\em exact functor} for the class of harmonic potentials can be
easily derived:
\begin{equation}
 V(q)=\frac{m\omega^2}2~q^2 ~~~~\longrightarrow~~~~~
 W(q) = \frac1{2\beta}\ln\frac{\sinh\beta\hbar\omega}{\beta\hbar\omega}
    + \frac{m\omega}{\beta\hbar}~\tanh\frac{\beta\hbar\omega}2~~q^2 ~.
\label{e.ho.Wexact}
\end{equation}
For a {\em linear} functor, this expression should be proportional to
$m\omega^2$: evidently this is true only in the classical limit,
$\beta\hbar\omega\,{\ll}\,1$ or $T\gg\hbar\omega$, where
Eq.~\eqref{e.ho.Wappr} is recovered. However, the mapping
$V\,{\to}\,W$ can surely be {\em locally} linear, namely
$V{+}\varepsilon{\delta}V\,{\to}\,W{+}\varepsilon\delta{W}$ with
$\delta{W}$ independent of the small parameter $\varepsilon$. Hence,
$W_{_{\rm{BPA}}}(q)$ is reliable only when the temperature overcomes
the typical quantum energy scale $\hbar\omega$; for instance, taking
$\omega^2\,{\sim}\,V''(q_{\rm{m}})/m$ ($q_{\rm{m}}$ being the minimum
of $V(q)$), a pair of hydrogen molecules has typically
$\hbar\omega\sim{10^2}$~K~\cite{VaiaT1990} and references therein]
and using $W_{_{\rm{BPA}}}(q)$ would only be reliable at very high
$T\,{\gg}\,10^2$~K, i.e., just in the classical limit. Such an
approximation is indeed used in the high-$T$ propagator of
path-integral Monte Carlo algorithms in order to improve convergence
in the Trotter number~\cite{TakahashiI1984}. Hence, the
approximation~(29) of BPA can ``reproduce quantum distributions''
just when these are almost classical.


On the other hand, the use of the {\em exact} effective {\em
pair-potential}, rather than that obtained from Eq.~(29) of BPA, is a
good starting point for treating a not too dense quantum fluid by
means of a classical-like simulation, as shown in the last section of
BPA and as noted by several authors (see, e.g.,
\cite{ThirumalaiHB1984}] and many references cited in BPA). At
variance with the procedure of BPA, based on the heavy calculation of
a ({\em locally}) `linear functor' at fixed $T$, it would be more
practical to directly obtain the exact pair-potential $W(q)$ for the
chosen $V(\hat{q})$, a task that can easily be carried out at any
$T$.


\end{document}